# Analysis of Power Swing Characteristics of Grid-Forming VSC System Considering the Current Limitation Mode

Yongxin Xiong, *Member, IEEE*, Heng Wu, *Senior Member*, *IEEE*, Yifei Li, *Student Member*, and Xiongfei Wang, *Fellow, IEEE*,

*Abstract*—**This paper investigates power swing characteristics of grid-forming voltage source converter (GFM-VSC) systems considering the current limitation mode in both non-inertial and inertial GFM-VSC systems. Following grid faults, non-inertial GFM-VSC systems can re-synchronize with the grid but may experience significant power swings driven by its control dynamics, while inertial GFM-VSC systems may exhibit loss of synchronization (LOS), characterized by the divergence of the output angle in the active power control loop. These behaviours are different from conventional synchronous generator (SG)-based systems, where power swings are typically characterized by physical angle deviations among power sources. Based on these findings, this paper explores the performance of traditional impedance-based swing detection schemes in GFM-VSC systems. The theoretical analysis is validated through various simulations using the PSCAD/EMTDC platform, covering both single and multi-machine system scenarios.**

*Index Terms*—**Voltage source converters, Droop-based grid-forming control, Circular current limiter, Power swing blocking, Out-of-step tripping**

## I. Introduction

THE growing utilization of power electronic converters introduces new control dynamics in power systems that differ from those of synchronous generators (SGs) [1], [2]. Although grid-following (GFL) control remains the predominant control scheme for voltage source converters (VSCs) across various applications, it faces challenges when connected to weak grids, mainly due to the adverse effects of the phase-locked loop (PLL) [3]. To mitigate these stability concerns, grid-forming (GFM) control has been developed and widely adopted across various applications [4]-[7]. With different implementations of active power control schemes, GFM-VSC systems can be categorized into non-inertial and inertial types, dependent on their capability to provide system inertia, as illustrated in [6]. Those different active power control schemes of GFM-VSCs introduce new power swing characteristics, necessitating a deeper reevaluation of the conventional power swing detection methods.

In conventional power systems, it is well understood that power swings are characterized by ***physical angle*** oscillations between different power sources, and the divergence of these ***physical angles*** indicates the loss of synchronization (LOS) [8]. In SG-based systems, the impedance trajectory can be employed to characterize ***physical angle dynamics*** and power swings [9]. Specifically, the power swing blocking (PSB) function is adopted to differentiate between system faults and power swings, preventing the unnecessary tripping of distance protection relays during stable power swings [10]-[14]. Additionally, out-of-step tripping (OST) is used to distinguish between stable and unstable power swings. When an unstable power swing is detected, the OST function activates, initiating appropriate breaker actions to mitigate the risk of LOS and cascading failures [15]-[17].

Notably, both conventional PSB and OST functions rely on the dynamic analysis and impedance measurements of SG-based systems. However, with the increasing prevalence of VSC-connected systems, the power swing characteristics are significantly influenced by the control dynamics of VSCs. In [8] and [9], it is indicated that integrating large-scale wind power can modify power swing dynamics, potentially causing maloperations for traditional PSB and OST functions. In [18], it is demonstrated that the fault response of VSC-dominated systems is highly dependent on VSC control dynamics, which in turn alters the impedance trajectory and presents challenges for the reliability of traditional impedance-based protections. In [19], the effects of different short-circuit characteristics of VSCs on PSB and OST functions are discussed. While pioneering works in this direction, these studies heavily rely on simulation-based observations, which are case-specific and lack comprehensive analytical insights without accounting for the control impacts of VSCs. To address those challenges, our previous work has demonstrated that the power swing characteristics in GFL-VSC systems are influenced by VSC control dynamics, and that LOS is characterized by the divergence of the ***control-dependent angle***, rather than the ***physical angle*** [5]. Nevertheless, the dynamics of GFM-VSC are different from GFL-VSC, necessitating a detailed investigation into their power swing characteristics and the implications for traditional PSB and OST functions.

In our previous work [4]-[7], the transient behaviour of the GFM-VSC has been analyzed based on large-signal nonlinear dynamic models, and it is found that the transient dynamic is governed by the ***control-dependent angle*** generated from the power-synchronization control (PSC) loop of GFM-VSC.

Nevertheless, the existing literature lacks a comprehensive

Yongxin Xiong is with the Hong Kong Polytechnic University, Hong Kong, China. (e-mail: yongxin.xiong@polyu.edu.hk)

Heng Wu, Yifei Li are with AAU Energy, Aalborg University, 9220 Aalborg, Denmark (e-mail: hew@energy.aau.dk; yili@energy.aau.dk).

Xiongfei Wang is with KTH Royal Institute of Technology, Stockholm, Sweden, (email: xiongfei@kth.se). (*Corresponding Author: Xiongfei Wang and Heng Wu*) analysis of the relationships between the ***control-dependent angle*** and power swings in GFM-VSC systems. Consequently, the effectiveness of the traditional impedance-based power swing detection method, which relies on detecting the ***physical angle***, remains an open issue. Given the fundamental differences in transient dynamics between SG and GFM-VSC systems, it is imperative to investigate the impact of GFM control on swing characteristics and to assess the efficacy of these traditional impedance-based detection methods.

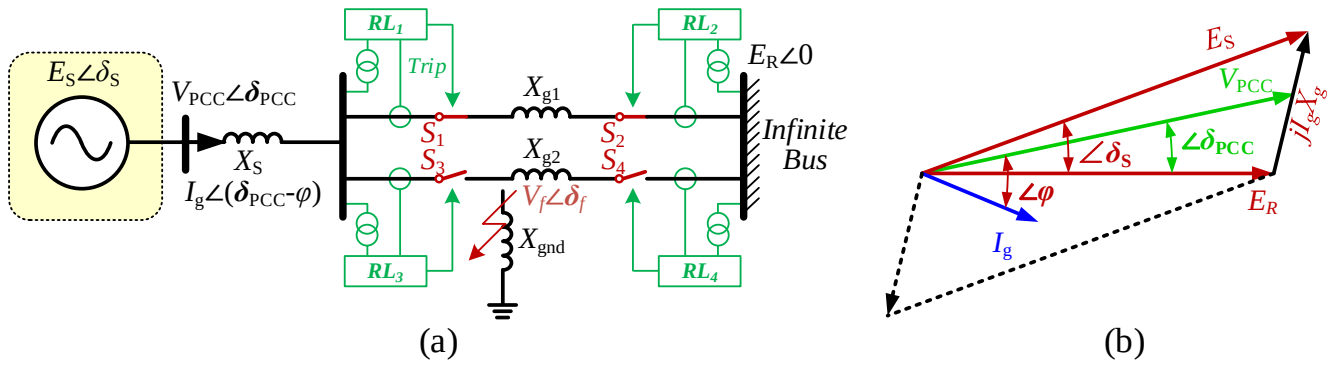

Fig. 1. Network and phasor diagram of SG-based power system: (a) network with parallel transmission lines, (b) phasor diagram.

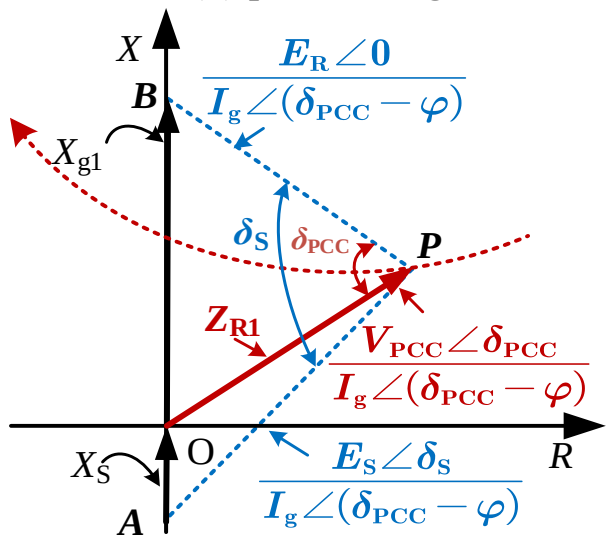

Fig. 2. Impedance trajectory of $Z_{R1}$ measured by $RL_1$ during post-fault power swings in SG-based systems.

Referring to our previous work in [4]-[7], impedance trajectories during power swings for GFM-VSC are studied in [20]. That study confirms that power swings in GFM-VSC systems behave differently from those in SG-based systems under current limitation conditions. However, merely noting that traditional impedance-based methods may not perform reliably in GFM-VSC systems is insufficient; it is equally critical to understand how the ***control-dependent angle*** fundamentally influences power swing dynamics.

Building on our previous work in [5], this paper conducts a theoretical analysis and physical insight into the impact of the ***control-dependent angle*** in GFM-VSC systems on power swing characteristics, considering the current limitation mode. The main contributions can be summarized as follows:

- In non-inertial GFM-VSC systems, the VSC system may suffer from significant power swings and enter the current limiting mode when the fault clearing time (FCT) exceeds the critical clearing time (CCT). During this process, the power swing is characterized by the ***control-dependent angle*** (i.e., $\delta_{PSC}$ from the PSC loop and current phase angle $\varphi$ generated from current limiting control), which is fundamentally distinct from SG swing dynamics driven by the ***physical angle***.
- In inertial GFM-VSC systems, the LOS may occur when the FCT exceeds the CCT. In this case, the GFM-VSC system may repeatedly switch between the voltage control mode and the current limitation mode, resulting in more complex and hybrid swing dynamics: it exhibits similar swing dynamics as that of non-inertial GFM-VSC under the current limiting mode while featuring similar swing dynamics as that of SG under the voltage control mode.
- It is demonstrated that the current phase angle $\varphi$ evolves as a function of $\delta_{PSC}$ under the current limitation mode. Consequently, despite the markedly different swing dynamics compared with SG systems, this relationship induces a leftward shift in the impedance trajectory of GFM-VSC in the complex plane as $\delta_{PSC}$ increases. Leveraging this insight, this paper, for the first time, reveals that the traditional impedance-based power swing detection methods for SG-based systems may still operate properly for GFM-VSC under weak grid conditions, but may malfunction as grid stiffness increases. To address this challenge, new design guidelines for setting the PSB and OST relays are proposed to ensure their effectiveness in GFM-VSC systems under various grid conditions.

The remainder of the paper is organized as follows: Section II introduces impedance-based power swing detection in SG-based systems. Sections III and IV analyze the transient of non-inertial and inertial GFM-VSC systems in the current limitation mode, respectively. Section V analyzes the control impact of GFM-VSC on power swings under the current limitation mode. Theoretical analyses are validated through simulations in the PSCAD/EMTDC platform in Sections VI and VII, with case studies in single-machine and multi-machine systems. Conclusions are drawn in Section VIII.

## II. Impedance-Based Power Swing Detection Method

The fundamental basis of power swing in an SG-based system is briefly reviewed in this section. For detailed knowledge of swing characteristics, please refer to [5]. Fig. 1 (a) illustrates the circuit diagram of the SG-based systems, whose phasor diagram is introduced in Fig. 1 (b) [13]. $E_S$, $V_{PCC}$, and $E_R$ are the internal voltage of SG, the PCC voltage, and the voltage of the infinite bus. The SG has a phase angle leading to the infinite bus (i.e. $\delta_S$). $\varphi$ represents the angle difference between $V_{PCC}$ and $I_g$. $X_S$, $X_{g1}$, $X_{g2}$, and $X_{gnd}$ indicate the internal impedance of the generator, the impedance of the transmission lines $L_1$ and $L_2$, and the grounding impedance, respectively. The close fault to the PCC on $L_2$ is considered for a clear explanation of the dynamic process of this SG-based system, and the power swing situations are observed after the fault clearance by open switches $S_3$ and $S_4$. The impedance measured by $RL_1$, i.e., $Z_{R1}$, is represented as (1) [5].

$$\begin{aligned} Z_{R1} &= \frac{V_{PCC}\angle\delta_{PCC}}{I_g\angle(\delta_{PCC}-\varphi)} \\ &= \left(jX_S + jX_{g1}\right)\cdot\frac{E_S\angle\delta_S}{E_S\angle\delta_S - E_R} - jX_S \end{aligned} \tag{1}$$

Referred to (1), the impedance trajectory during post-fault power swings can be plotted in Fig. 2. The vector ***OP*** denotes the measured $Z_{R1}$, while the vectors ***AP*** and ***BP*** represent the $E_S\angle\delta_S/I_g\angle(\delta_{PCC}-\varphi)$ and $E_R\angle 0/I_g\angle(\delta_{PCC}-\varphi)$, respectively. The angle difference between the ***AP*** and ***BP*** is equal to $\delta_S$. Thus, $Z_{R1}$ could be used to predict the variations of $\delta_S$, and then characterize the post-fault power swing [13]. The impedance locus may travel towards the left plane if the angle $\delta_S$ increases, indicating a higher risk of LOS. Therefore, in the conventional SG-based system, it is the ***physical angle*** $\delta_S$ that characterizes the power swing, and the swing trajectory can be monitored by the moving trajectory of $Z_{R1}$. Hence, tracking the impedance trajectory of $Z_{R1}$ is one of the most practical methods to detect power swing situations [12].

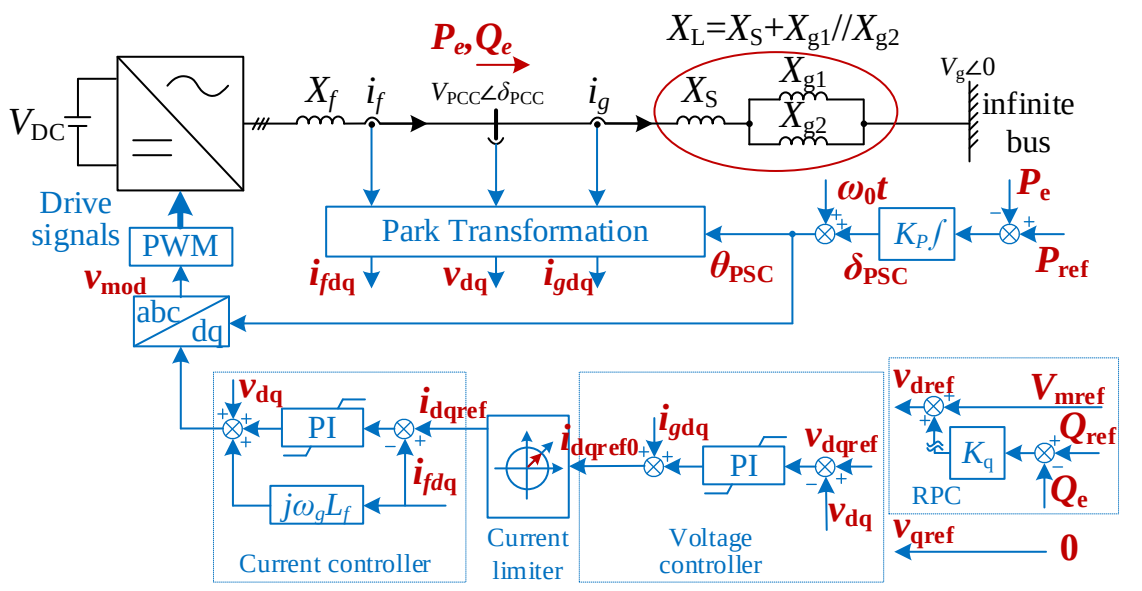

Fig. 3. Control diagram of the non-inertial GFM-VSC system.

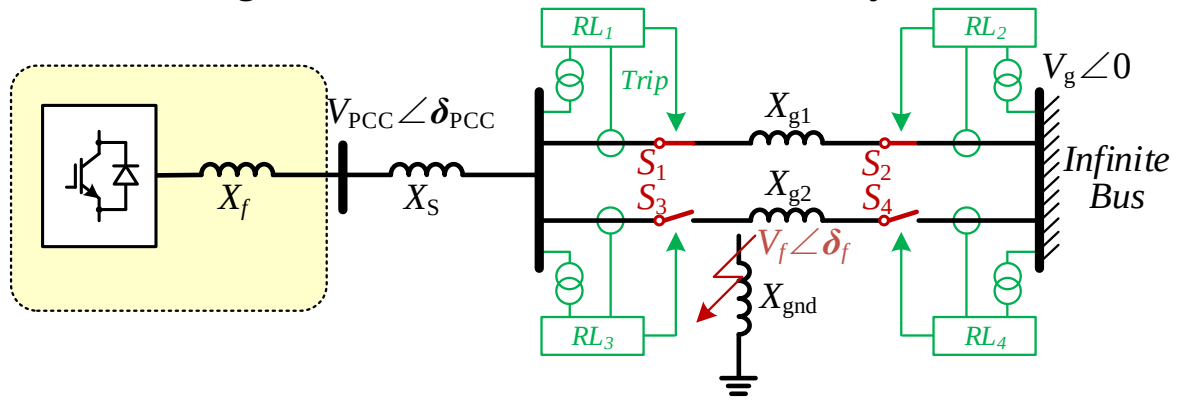

Fig. 4. Single-line diagram of the GFM-VSC system.

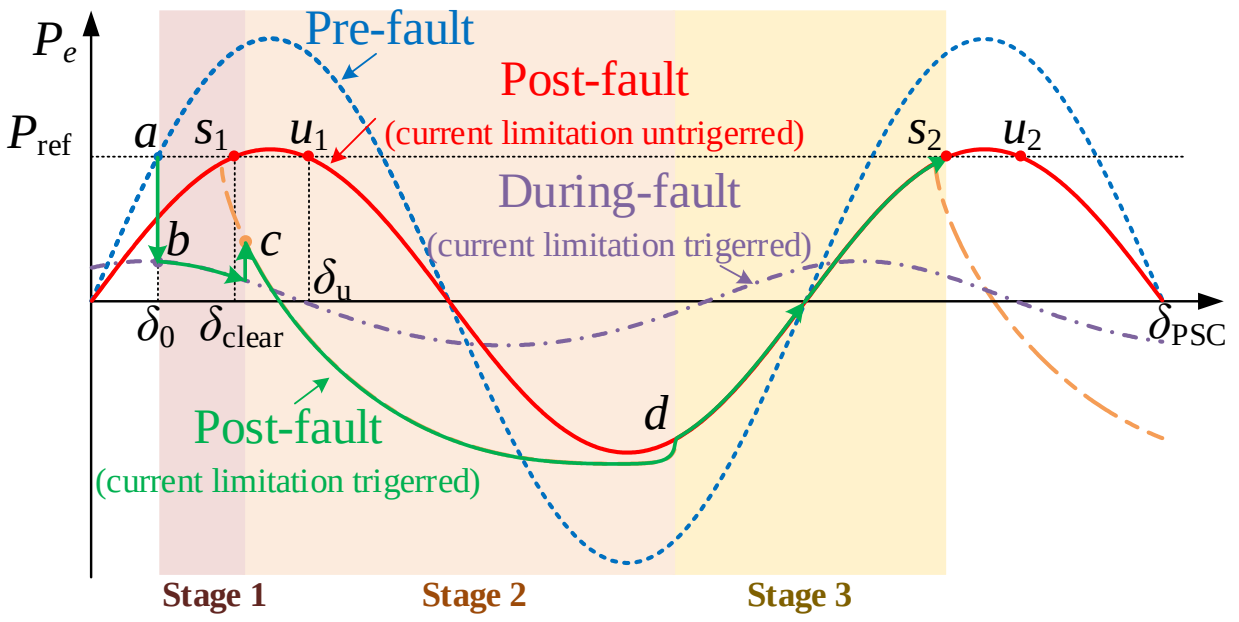

Fig. 5. $P$-$\delta$ curves during recovery mode.

## III. Transient Analysis of Non-inertial GFM-VSC Systems

### *A. System Description*

The power control of GFM-VSC systems can be classified as non-inertial (without inertia response) and inertial (with inertia response) systems [6]. Fig. 3 illustrates the control diagram of a non-inertial GFM-VSC system with the circular current limiter. As a preliminary analysis, a constant DC-link voltage ($V_{DC}$) is assumed for the GFM-VSC system [4]. The converter is connected to the grid through an $L$ filter, with $X_f$ being its impedance inductance. The *Power-frequency* (*P-f*) droop control is employed, and the voltage reference is given by the inner dual-loop proportional-integral (PI) control. The GFM-VSC interfaces with the infinite bus through two paralleled transmission lines, $X_{g1}$ and $X_{g2}$, represented by its inductance [21]. $V_{PCC}\angle\delta_{PCC}$ and $V_g\angle 0$ denote the voltage of the PCC and the infinite bus, respectively. $i_f$ and $i_g$ represent the converter-side and the grid-side current, respectively. $X_S$ and $X_f$ represent the equivalent impedance of the transformer and the converter output filter, respectively. An inner voltage loop is used to regulate the voltage at the PCC to track its reference.

In Fig. 3, $\delta_{PSC}$ denotes the phase angle generated by the PSC loop and respective to the infinite bus, which can be given by:

$$\begin{aligned}\delta_{PSC} &= \theta_{PSC} - \theta_g = K_p \int (P_{ref} - P_e) + \omega_0 t - \omega_0 t \\ &= K_p \int (P_{ref} - P_e)\end{aligned} \tag{2}$$

where $P_{ref}$ and $P_e$ represent the reference and measurement of the active power for power synchronizing control. $\omega_0$ denotes the grid frequency, and $K_p$ is the integral gain. $\theta_g$ denotes the angle of the infinite bus (0 degrees). The reactive power control is employed by a *Q-V* droop control presented as [22]:

$$v_{dref} = V_{mref} + K_q \cdot (Q_{ref} - Q_e) \tag{3}$$

where $V_{mref}$ and $v_{dref}$ represent the initial and revised *d-axis* voltage reference, and $Q_{ref}$ and $Q_e$ represent the reactive power reference and measurement. $K_q$ is the *Q-V* droop gain for RPC. $v_{qref}$ in Fig. 3 is set to 0, referred to [4]-[7].

### *B. Circular Current Limiter*

The circular current limiter illustrated in [7] is utilized in this study. For the non-inertial GFM-VSC system operating under the current limitation mode, the current can be regulated to $I_M\angle\delta_{PSC}+\varphi$, where $I_M$ represents the maximum allowable converter-side current magnitude, and $\varphi$ denotes the current phase angle (defined by $\tan\varphi = i_{gq}/i_{gd}$). As referenced in our previous study [7], the current references can be expressed as:

$$i_{dqref} = k_{Pi}\left(v_{dqref} - v_{dq}\right) + i_{gdq} \tag{4}$$

where $k_{Pi}$ is the voltage control proportional gain. $i_{dqref}$ and $v_{dqref}$ represent the current and voltage references on *dq-axis*, while $i_{dqref}$ and $v_{dqr}$ represent the measured current and voltage. Notice that $i_{fdq}=i_{gdq}$ since the dynamic of the capacitor can be ignored [7]. Further, with $i_{fdq}= i_{dqref}$, (4) can be expressed as

$$\begin{aligned}& i_{fdq} = \sigma i_{dqref0} = \sigma k_{Pi}\left(v_{dqref} - v_{dq}\right) + \sigma i_{gdq} \\ & \Rightarrow R_e i_{fdq} = v_{dqref} - v_{dq} \quad \left(R_e = \frac{1-\sigma}{\sigma k_{Pi}}\right)\end{aligned} \tag{5}$$

where $\sigma$ is a positive value, i.e., only the magnitude of current is scaled down when the circular current limiter is triggered, $R_e$ is a positive variable and can be represented as [7]:

$$R_e = \max\left\{0,\ \mathrm{Re}\left\{\sqrt{\frac{V_{mref}^2 - 2V_{mref}V_g\cos\delta_{PSC} + V_g^2}{I_M^2} - X_L^2}\right\}\right\} \tag{6}$$

When the current limitation is activated, the GFM-VSC with the circular current limiter can be modelled as a voltage source behind an adjustable resistance $R_e$, as described in (6). $X_L$ represents the equivalent grid line impedance, i.e., $X_L=X_S+X_{g1}//X_{g2}//X_{gnd}$ during the fault, and $X_L=X_S+X_{g1}$ post-fault. $R_e$=0 when the current limitation is not triggered [7].

### *C. Power Swing under Current limitation mode after Fault*

Referring to (2), the *power-angle* curve can be plotted as the blue dash line in Fig. 5 before the fault, with the GFM-VSC system initially operating at ***Point a***, and the physical angle $\delta_{PCC}$ is equal to the control-dependent angle $\delta_{PSC}$. Once the fault occurs and the FCT is larger than the CCT, the non-inertial GFM-VSC system can experience significant power swings, which can be divided into three stages:

*1) Stage 1: During Fault*

During the fault, the equivalent circuit during a fault can be simplified as Fig. 6 (a). The green line part is the equivalent grid impedance and voltage source during the fault. The electric power of GFM-VSC can be further expressed as (7):

$$P_e = \frac{R_e\left(V_{PCC}^2 - V_{PCC}V_{ge}\cos\delta_{PSC}\right)}{R_e^2 + X_L^2} + \frac{X_L V_{PCC} V_{ge}}{R_e^2 + X_L^2}\sin\delta_{PSC} - I_M^2 R_e$$
$$\left(X_L = X_S + X_{g1}//X_{g2}//X_{gnd},\ V_{ge} = \frac{X_{gnd}}{X_{g1}//X_{g2}//X_{gnd}}V_g\right) \quad (7)$$

where $X_L$ and $V_{ge}$ represent the equivalent grid impedance and voltage magnitude, respectively. Consequently, the *power-angle* curve can be plotted with the purple dash line in Fig. 5. The operating point shifts immediately from ***Point a*** to ***Point b*** and then moves along the purple power-angle curve.

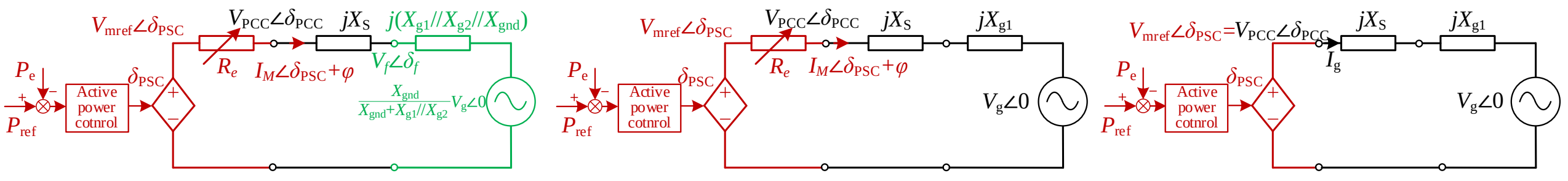

Fig. 6. Equviliant circuit of the non-inertial GFM-VSC system: (a) Stage 1: during fault (current limitation mode), (b) Stage 2: after fault (current limitation mode), (c) Stage 3: after fault (voltage control mode).

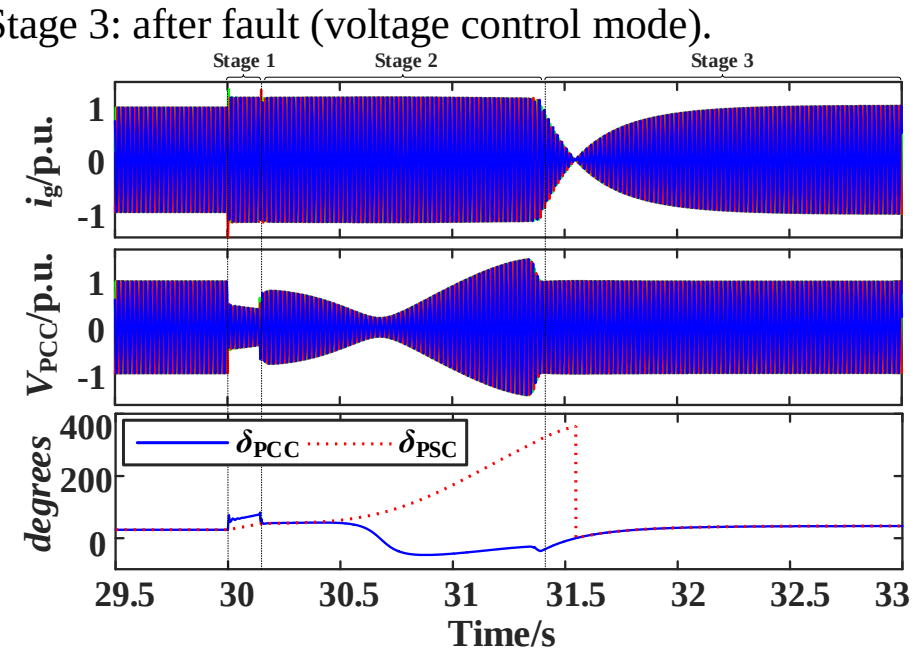

Fig. 7 Simulation results in non-inertial GFM-VSC system.

TABLE I
MAIN PARAMETERS OF BOTH THE SG AND THE VSC SYSTEMS

| Symbol | Descriptions | Values |
|---|---|---|
| $V_g$ | The L-L AC voltage rating | 20 kV (1 p.u.) |
| $V_{mref}$ | Voltage reference for reactive loop | 20 kV (1 p.u.) |
| $P_{base}$ | The power rating of the VSC | 250 MW (1 p.u.) |
| $Z_{base}$ | Rated impedance | 1.6 Ω (1 p.u.) |
| $X_S$ | Equivalent impedance of transformers | 0.30 p.u. |
| $X_f$ | Equivalent impedance of the filter | 0.20 p.u. |
| $X_{g1}$ | Equivalent impedance of Line 1 | 0.35 p.u. |
| $X_{g2}$ | Equivalent impedance of Line 2 | 0.35 p.u. |
| $X_{gnd}$ | Equivalent impedance of the fault | 0.05 p.u. |

TABLE II
MAIN CONTROL PARAMETERS OF THE NON-INERTIAL GFM-VSC SYSTEM

| Parameters | Values | Parameters | Values |
|---|---|---|---|
| $K_P$ (p.u.) | 0.01 | $K_q$ (p.u.) | 0.05 |
| $P_{ref}$ (p.u.) | 1.00 | $Q_{ref}$ (p.u.) | 0.00 |
| CCT (s) | 0.13s | FCT (s) | 0.14s |

*2) Stage 2: After Fault with Current Limitation Mode*

When the FCT exceeds the CCT and the fault is cleared by opening $S_3$ and $S_4$ in Fig. 4, the operating point may move to ***Point c***. After the fault clearance, the ***physical angle*** $\delta_{PCC}$ may still be large, and the system can not immediately quit the current limitation mode. In this case, $\delta_{PCC} \neq \delta_{PSC}$. Consequently, (7) can be rewritten as:

$$P_e = \frac{R_e\left(V_{PCC}^2 - V_{PCC}V_g\cos\delta_{PSC}\right)}{R_e^2 + X_L^2} + \frac{X_L V_{PCC} V_g}{R_e^2 + X_L^2}\sin\delta_{PSC} - I_M^2 R_e \quad (8)$$
$$(X_L = X_S + X_{g1})$$

where $X_L = X_S + X_{g1}$. Referring to (8), the power-angle curve is plotted as shown in Fig. 5. The equivalent circuit can be represented as shown in Fig. 6 (b). After the fault is cleared, the power reference ($P_{ref}$) remains greater than the electrical power ($P_e$), causing $\delta_{PSC}$ to continue to increase, as described by (2). During this period, the operating point moves from ***Point c*** to ***Point d***, and the non-inertial GFM-VSC system remains in the current limitation mode for a certain period. As $\delta_{PSC}$ increases, significant power swings can occur.

*3) Stage 3: Existing from Current Limitation Mode*

The non-inertial GFM-VSC system may exit the current limitation mode at ***Point d***. The equivalent circuit after exiting is shown in Fig. 6 (c). During this process, $\delta_{PCC}$ is equal to $\delta_{PSC}$ again, The system may move with the red power-angle curve to ***Point s₂*** and stabilize at this point. The red power-angle curve in Fig. 5 can be expressed as:

$$P_e = \frac{V_{PCC}V_g}{X_S + X_{g1}}\sin\delta_{PSC} \quad (9)$$

To further illustrate these novel power swing characteristics, a case study is conducted using the system shown in Fig. 4, with key parameters provided in Tables I and II. In this case, a three-phase-ground fault occurs at $t$=30s, The fault is cleared by opening switches $S_3$ and $S_4$ after 0.14s, larger than the CCT, i.e., 0.13s. As a result, a significant power swing can be observed after fault clearance. The simulation results are presented in Fig. 7, it can be found that, in Stage 2, during the current limitation mode after fault clearance, the ***control-dependant angle*** ($\delta_{PSC}$) is not equal to the ***physical angle*** ($\delta_{PCC}$), $\delta_{PSC}$ increases larger than 360°, while $\delta_{PCC}$ only varies between -55° and 50°. Furthermore, the current $i_g$ is maintained at its limitation during this period. When the GFM-VSC quits the current limitation mode in Stage 3, the current $i_g$ may decrease to 0. This is a novel power swing characteristic compared with SG-based systems, in which $\delta_{PCC}$ may change significantly during the power swing.

## IV. TRANSIENT ANALYSIS OF INERTIAL GFM-VSC SYSTEMS

### *A. System Description*

In inertial GFM-VSC systems, the low-pass filter (LPF) can be employed to emulate the inertia response [6]. Fig. 8 illustrates the benchmark of an inertial GFM-VSC system with the circular current limiter. According to [7], the active and reactive power control loop can be represented as:

$$\begin{cases} \delta_{PSC} = \dfrac{K_p}{s}\cdot\dfrac{\omega_p}{s+\omega_p}\cdot\left(P_{ref} - \dfrac{3}{2}\cdot\dfrac{EV\sin\delta_{PSC}}{X_{g1}+X_S}\right) \\ V = V_{ref} + K_q\cdot\dfrac{\omega_q}{s+\omega_q}\cdot\left(Q_{ref} - \dfrac{3}{2}\cdot\dfrac{V^2 - EV\cos\delta_{PSC}}{X_{g1}+X_S}\right) \end{cases} \quad (10)$$

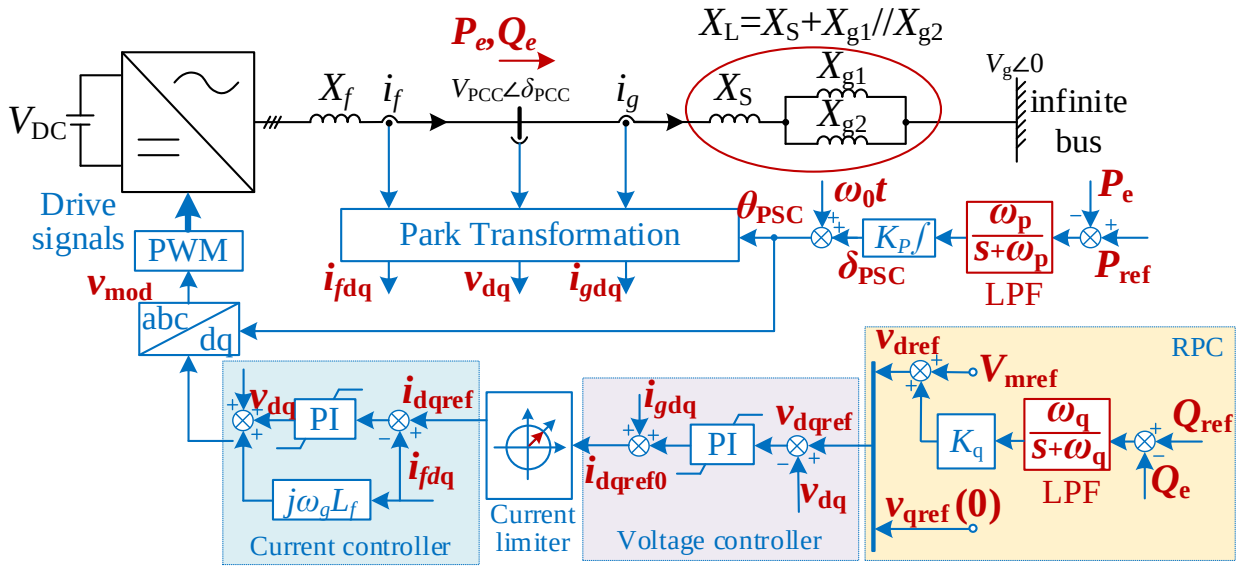

Fig. 8. Control diagram of the inertial GFM-VSC system.

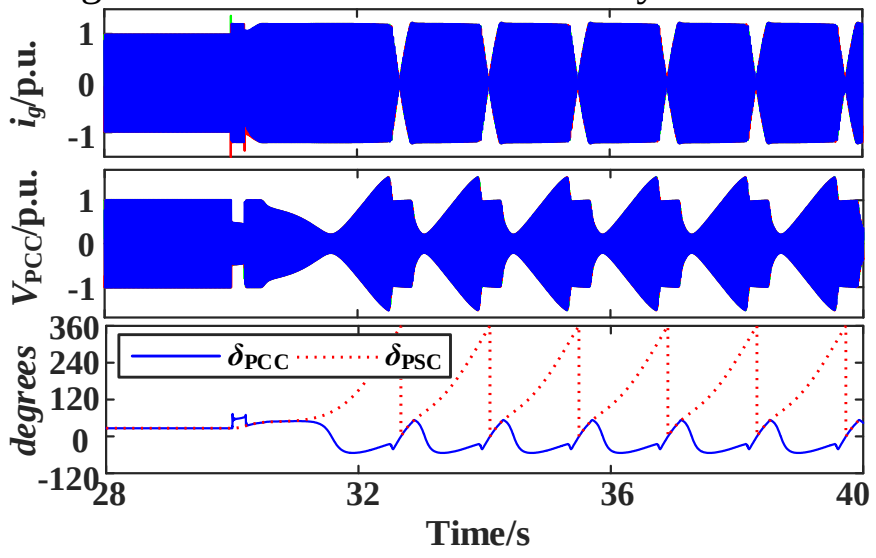

Fig. 9 Simulation results in inertial GFM-VSC systems with LOS.

TABLE III

MAIN CONTROL PARAMETERS OF THE INERTIAL GFM-VSC SYSTEM

| Parameters | Values | Parameters | Values |
|---|---|---|---|
| $K_P$ (p.u.) | 0.01 | $K_q$ (p.u.) | 0.05 |
| $\omega_P$ | 0.53×2π | $\omega_q$ | 0.53×2π |
| CCT (s) | 0.25s | FCT (s) | 0.28s |

where $P_{\text{ref}}$, $Q_{\text{ref}}$, and $V_{\text{ref}}$ are the reference values of active, reactive power, and voltage, respectively. $K_p$ and $K_q$ represent gains of the power synchronizing control loop and the $Q$-$V$ droop control loop. $\omega_P$ and $\omega_q$ are the cut-off frequencies in the LPFs [6]. Key parameters are referred to in Tables I and III.

### *B. Power Swing under Current Limitation Mode after Fault*

In this scenario, the FCT for the three-phase-ground fault is set to 0.28s, larger than the CCT (0.25s), and the simulation results are shown in Fig. 9. In this case, the system may experience LOS, and the system may repeatedly enter and exit the current limitation mode. It can also be found that the ***control-dependent angle*** $\delta_{\text{PSC}}$ diverges during the LOS, and the system alternates between the current limitation mode and voltage control mode. However, the ***physical angle*** $\delta_{\text{PCC}}$ oscillates between -55° and 54°. This is entirely different from the LOS in the SG-based system, where the LOS is characterized by the divergence of the ***physical angle***.

## V. Control Impact of GFM-VSC System with Current Limitation on Power Swing Characteristics

This section introduces the principles of the conventional impedance-based power swing detection method, and analyzes the impact of the GFM-VSC control with the circular current limitation mode on traditional power swing detection. It also reveals the physical insight into how the ***control-dependent angle*** characterizes power swings during current limitations.

### *A. Settings of Impedance-based PSB and OST Functions*

Since the benchmark systems of SG and GFM-VSC, as depicted in Fig. 1, Fig. 4 and Fig. 8, have identical grid connections and line parameters, to realize a fair comparison, the PSB and OST parameters, originally based on SG systems, are employed for both benchmark systems according to [8] and [13]. As the power swing characteristics of GFM-VSC

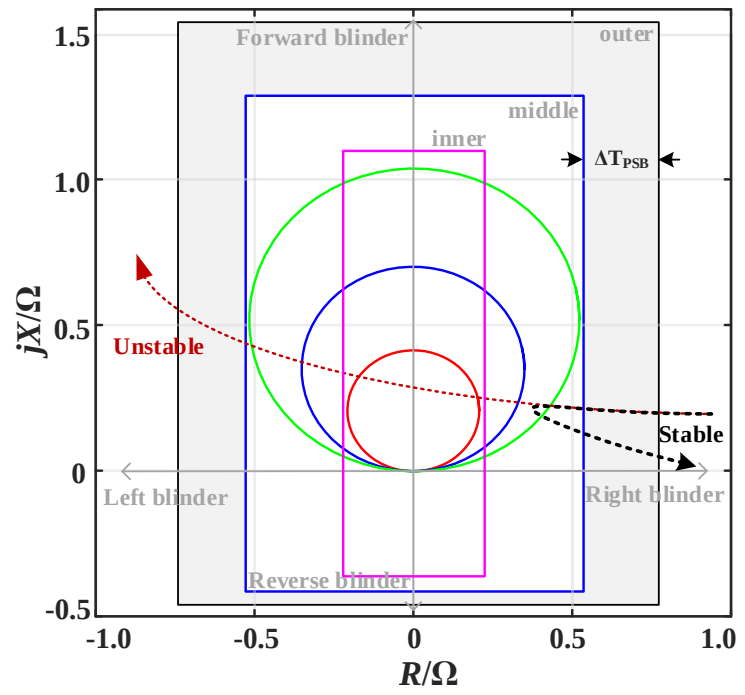

Fig. 10. Sketch diagram of impedance-based power swing detection method.

TABLE IV

SETTINGS OF AND DESCRIPTIONS OF RELAY RL1

| Settings | Description | Values |
|---|---|---|
| $Z_{L1}$ | The positive sequence of Line 1 | $j$0.560Ω |
| Outer forward reach | Outside the most reach of distance protection blocked by PSB | $j$1.583 Ω |
| Middle forward reach | Less than outer-forward blinder | $j$1.384 Ω |
| Inner forward reach | Less than middle-forward blinder | $j$1.185 Ω |
| Outer reserve reach | Outside maximum load impedance | - $j$0.495 Ω |
| Middle reserve reach | Less than the outer reverse blinder | - $j$0.447 Ω |
| Inner reserve reach | Less than the middle reverse blinder | - $j$0.393 Ω |
| Outer right reach | Less than maximum load impedance | 0.77 Ω |
| Outer left reach | Referred to the outer right blinder | -0.77 Ω |
| Middle right reach | Outside the most reach of distance protection blocked by PSB | 0.571 Ω |
| Middle left reach | Referred to the outer middle blinder | -0.571 Ω |
| Inner right reach | To ensure only unstable power swings can travel through it | 0.23 Ω |
| Inner left reach | Referred to the inner right blinder | -0.23 Ω |
| PSB time reach | Set to ensure the fastest power swing can be detected | 1.5 cycles |
| PSB reset delay | After that time PSB will be reset | 10 cycles |

systems (both non-inertial and inertial) under the voltage control mode after fault are similar to SG-based systems, this paper will focus on the scenarios under the current limitations. Referring to [8] and [11], the detailed settings of different blinders for power swing detection are shown in Table IV.

*1) PSB:* According to [8], the PSB function initiates timing when the impedance locus crosses the outer blinder, and it monitors the duration as the locus passes through the middle blinders. By comparing this interval to a reference time, $\Delta T_{\text{PSB}}$, the PSB function can distinguish between a system fault and a power swing. If the interval is shorter than $\Delta T_{\text{PSB}}$, the PSB identifies it as a fault and would not be activated. Conversely, if the interval is longer, the PSB detects a power swing, and would be activated and block related distance protections.

It indicates in [8] that the outer reverse blinder should be set below the absolute imaginary part of the maximum anticipated load impedance, and the outer forward blinder should be configured to cover the largest distance protection zone that is to be blocked by PSB. Based on these principles, the outer reverse blinder is set to 0.495Ω, given that the maximum load impedance is approximately 3.14-$j$0.33Ω. The outer forward blinder is set to 1.583Ω to encompass the Zone 3 distance protection. As there are no universally recognized guidelines for setting forward and reverse blinders, their impact on power swing detection is also examined and discussed in case studies.

*2) OST:* The OST function activates during unstable swing conditions to prevent LOS. As mentioned in [8] and [15], the OST function is triggered when the impedance trajectory crosses the inner blinder. To ensure robust performance, the inner blinder's position should be determined through stability analysis, allowing only unstable power swings to cross it.

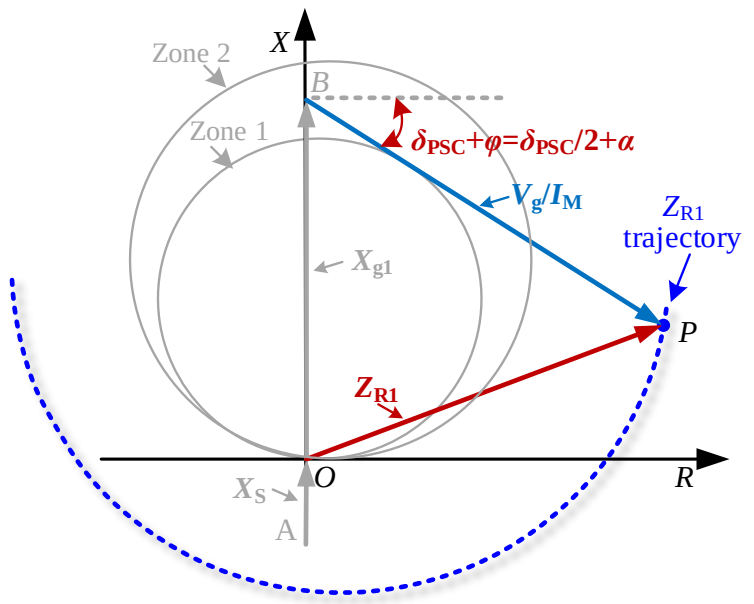

Fig. 11. Impedance trajectory with the current limitation triggered.

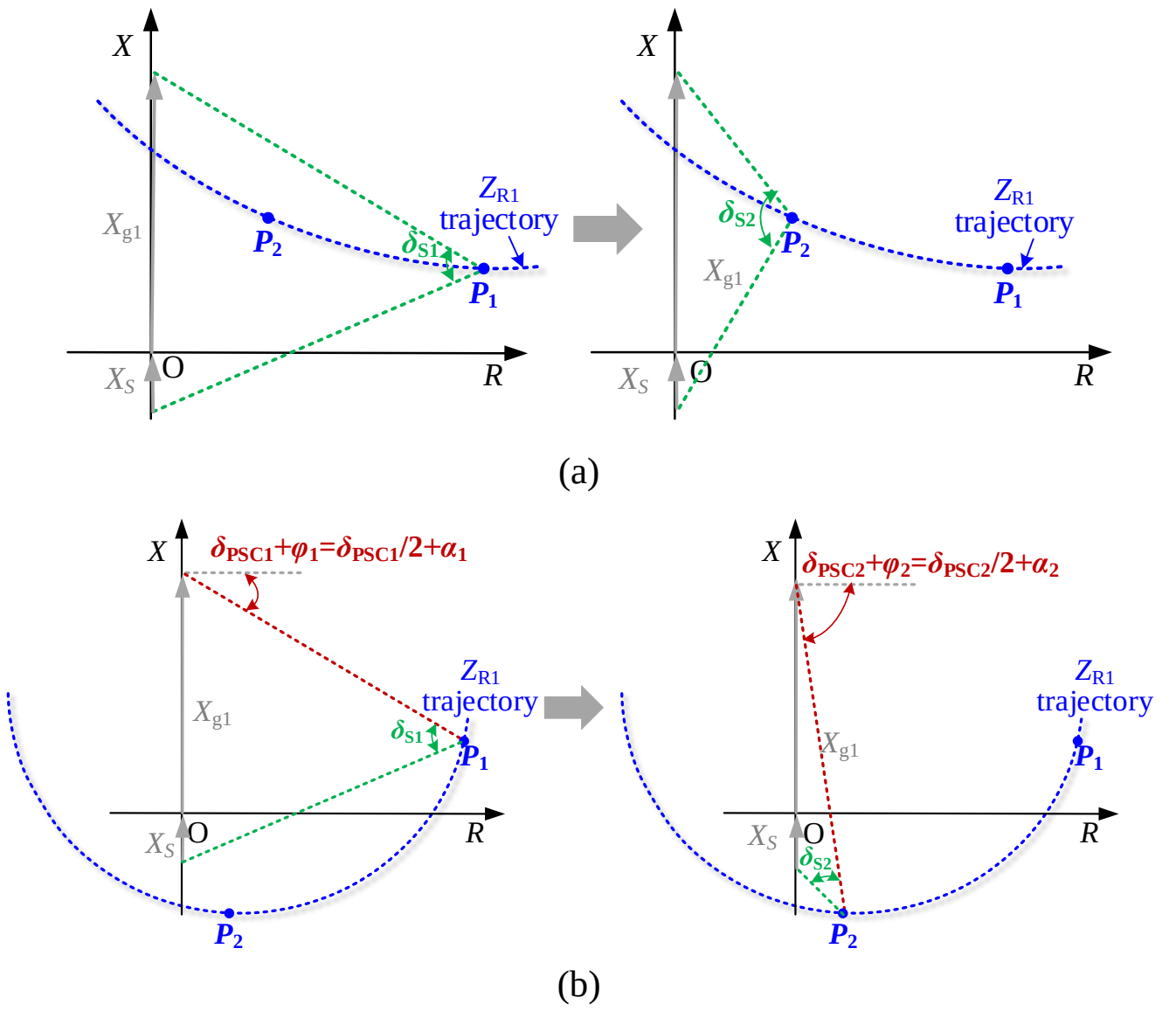

Fig. 12. Impedance trajectory predicted by different angles. (a) conventional SG-based systems, (b) non-inertial GFM-VSC under current limitation mode.

*B. Power Swing Characteristic in Non-inertial GFM-VSC*

As mentioned earlier, the non-inertial GFM-VSC can still synchronize with the power grid even if the FCT exceeds the CCT. However, during this period, the system may experience a significant power swing and operate under the current limitation mode. During the current limitation mode, the impedance measured by $RL_1$ can be expressed as:

$$Z_{R1} = \frac{V_{R1}\angle\delta_{R1}}{I_M\angle(\delta_{PSC}+\varphi)} = \frac{V_g + I_M\angle(\delta_{PSC}+\varphi)X_{g1}\angle\theta_{g1}}{I_M\angle(\delta_{PSC}+\varphi)} = \frac{V_g}{I_M}\angle-(\delta_{PSC}+\varphi)+X_{g1}\angle\theta_{g1} \tag{11}$$

where $\varphi$ represents the current phase angle, represented as:

$$\varphi = \arctan\left(\frac{i_{gq}}{i_{gd}}\right) \tag{12}$$

In (11), $V_g$ remains constant during the power swing, and $I_M$ represents the magnitude of the maximum current, which also remains constant with the current limitation mode. With $V_g$, $I_M$, and $X_{g1}\angle\theta_{g1}$ remaining constant, $Z_{R1}$ is dependent on the dynamics of $\delta_{PSC}+\varphi$, forming a circle path in the complex plane with a radius of $V_g/I_M$, as shown in Fig. 11. With $\delta_{PSC}+\varphi$ increasing, the $Z_{R1}$ trajectory moves clockwise towards the left half-plane. To further illustrate the impact of the ***control-dependent angle*** on the impedance trajectory, the output current of non-inertial GFM-VSC systems during the current limitation mode can be expressed as:

$$I_M\angle(\delta_{PSC}+\varphi) = \frac{V_{mref}\angle\delta_{PSC} - V_g\angle 0}{R_e + jX_L} \tag{13}$$

In (13), $X_L$ represent the total system impedance as shown in Fig. 6 (c) and defined in (8), as we only focus on the post-fault power swing. $R_e$ is a positive, variable virtual resistance, as defined in (6). By calculating the real and imaginary parts of the current, the tangent value of $\delta_{PSC}+\varphi$ can be derived as:

$$\tan(\delta_{PSC}+\varphi) = \frac{V_{mref}R_e\sin\delta_{PSC} - V_{mref}X_L\cos\delta_{PSC} + V_g X_L}{V_{mref}R_e\cos\delta_{PSC} + V_{mref}X_L\sin\delta_{PSC} - V_g R_e} \tag{14}$$

To simplify (14), a variable virtual angle $\alpha$ is introduced as:

$$\begin{cases} \sin\alpha = R_e\big/\sqrt{R_e + X_L} \\ \cos\alpha = X_L\big/\sqrt{R_e + X_L} \end{cases} \tag{15}$$

As aforementioned, $V_{mref}$ and $V_g$ are constant and set to 1.0 p.u., when substituting (15) into (14), (14) can be simplified as (16), and $\delta_{PSC}+\varphi$ is equal to $\delta_{PSC}/2+\alpha$. Since $\alpha$ is the only term related to $\delta_{PSC}$, as indicated in (6) and (15), the impedance trajectory is solely characterized by $\delta_{PSC}$. Given that $R_e$ is a positive value, $\alpha$ is always positive according to (6) and (15), therefore, the sum $\delta_{PSC}/2+\alpha$ is invariably larger than $\delta_{PSC}/2$. As $\delta_{PSC}$ increases, the $Z_{R1}$ trajectory rotates clockwise toward the left half-plane. Based on (11), the impedance trajectory under the current limitation mode is illustrated in Fig. 11.

A graphical comparison of impedance trajectories between SG and the non-inertial GFM-VSC is presented in Fig. 12, illustrating the movement from point $P_1$ to $P_2$, from which, fundamental differences can be identified:

- ***Conventional SG-based systems:*** In traditional SG-based systems, the internal voltage magnitude of SG (i.e., $E_S$) remains relatively constant during a power swing, causing the impedance trajectory of SG to move leftwards as $\delta_S$ increases according to (1). As depicted in Fig. 12 (a), the impedance trajectory moves from operating point $P_1$ to $P_2$ as $\delta_S$ increases from $\delta_{S1}$ to $\delta_{S2}$.
- ***Non-inertial GFM-VSC system with current limitation mode:*** In contrast, in the non-inertial GFM-VSC system with current limitation mode, $V_{PCC}$ may have a significant change to maintain the constant magnitude of the output current ($i_g$). Therefore, the leftward movement of the impedance trajectory does not necessarily indicate an increase of $\delta_S$ in such systems. As depicted in Fig. 12 (b), the impedance trajectory in the GFM-VSC system also moves leftwards from operating point $P_1$ to $P_2$, yet the movement is characterized by the ***control-dependent angle*** $\delta_{PSC}/2+\alpha$, which is solely dependent on $\delta_{PSC}$.

*C. Power Swing Characteristic in Inertial GFM-VSC*

In an inertial GFM-VSC system, unlike a non-inertial GFM-VSC system, the LOS can occur if the FCT exceeds the CCT. During the LOS period, the current limitation mode can be triggered repeatedly, causing the GFM-VSC system to

$$\begin{aligned}\tan\left(\delta_{\mathrm{PSC}}+\varphi\right)&=\frac{\sin\delta_{\mathrm{PSC}}\sin\alpha-\cos\delta_{\mathrm{PSC}}\cos\alpha+\cos\alpha}{\cos\delta_{\mathrm{PSC}}\sin\alpha+\sin\delta_{\mathrm{PSC}}\cos\alpha-\sin\alpha}=\frac{\cos(\delta_{\mathrm{PSC}}+\alpha)-\cos\alpha}{\sin\alpha-\sin(\delta_{\mathrm{PSC}}+\alpha)}\\&=\frac{\cos\left[\left(\delta_{\mathrm{PSC}}+\alpha+\alpha\right)/2+\left(\delta_{\mathrm{PSC}}+\alpha-\alpha\right)/2\right]-\cos\left[\left(\delta_{\mathrm{PSC}}+\alpha+\alpha\right)/2-\left(\delta_{\mathrm{PSC}}+\alpha-\alpha\right)/2\right]}{\sin\left[\left(\delta_{\mathrm{PSC}}+\alpha+\alpha\right)/2-\left(\delta_{\mathrm{PSC}}+\alpha-\alpha\right)/2\right]-\sin\left[\left(\delta_{\mathrm{PSC}}+\alpha+\alpha\right)/2+\left(\delta_{\mathrm{PSC}}+\alpha-\alpha\right)/2\right]}\\&=\frac{-2\sin\left[\left(\delta_{\mathrm{PSC}}+\alpha+\alpha\right)/2\right]\cdot\sin\left[\left(\delta_{\mathrm{PSC}}+\alpha-\alpha\right)/2\right]}{-2\cos\left[\left(\delta_{\mathrm{PSC}}+\alpha+\alpha\right)/2\right]\cdot\sin\left[\left(\delta_{\mathrm{PSC}}+\alpha-\alpha\right)/2\right]}=\tan\left(\delta_{\mathrm{PSC}}/2+\alpha\right)\end{aligned}\tag{16}$$

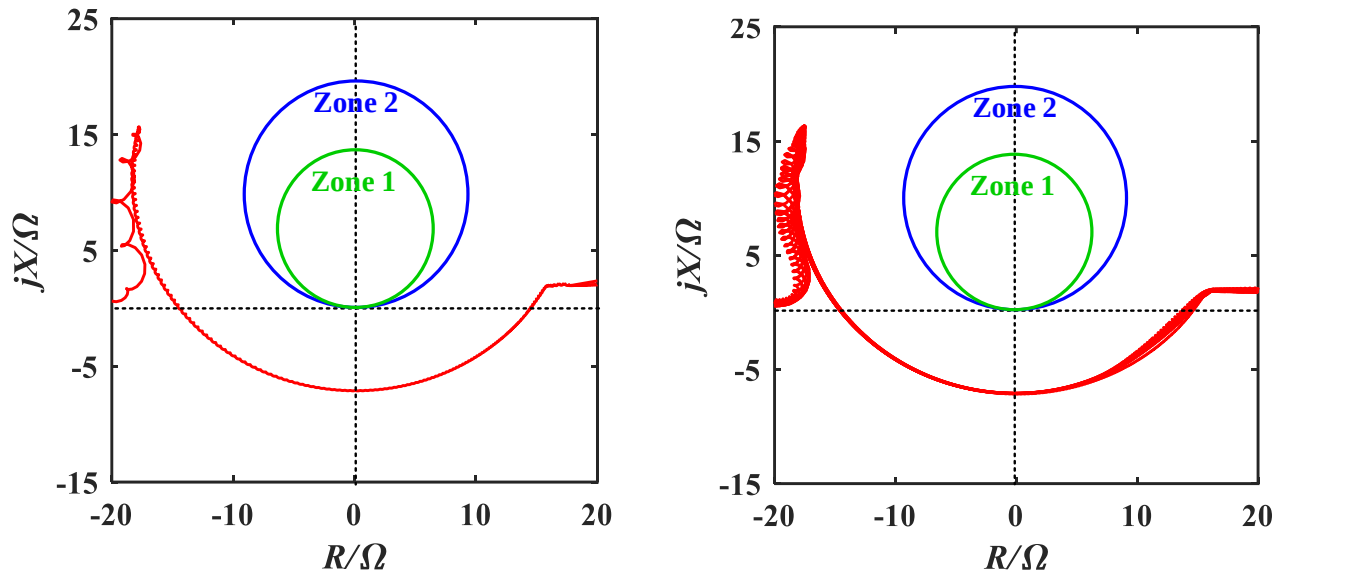

(a) (b)

Fig. 13. Impedance trajectories of: a) the power swing in a non-inertial GFM-VSC system, b) the LOS in an inertial GFM-VSC system.

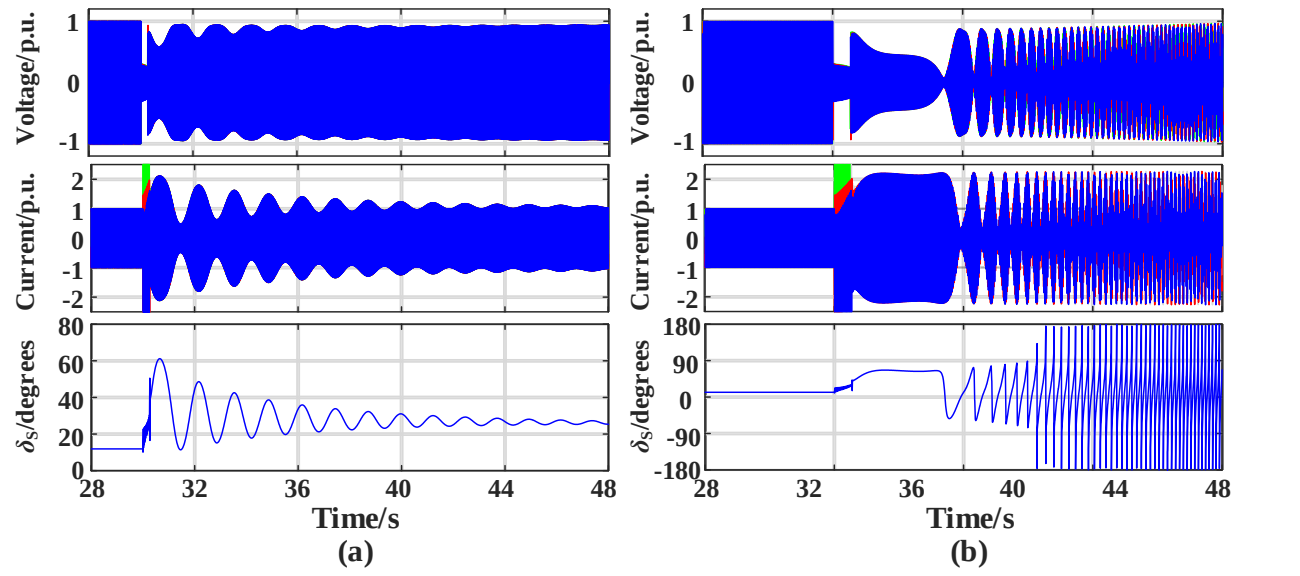

Fig. 14. Simulation results with different CCT in the SG-based system: (a) stable power swing, (b) unstable power swing.

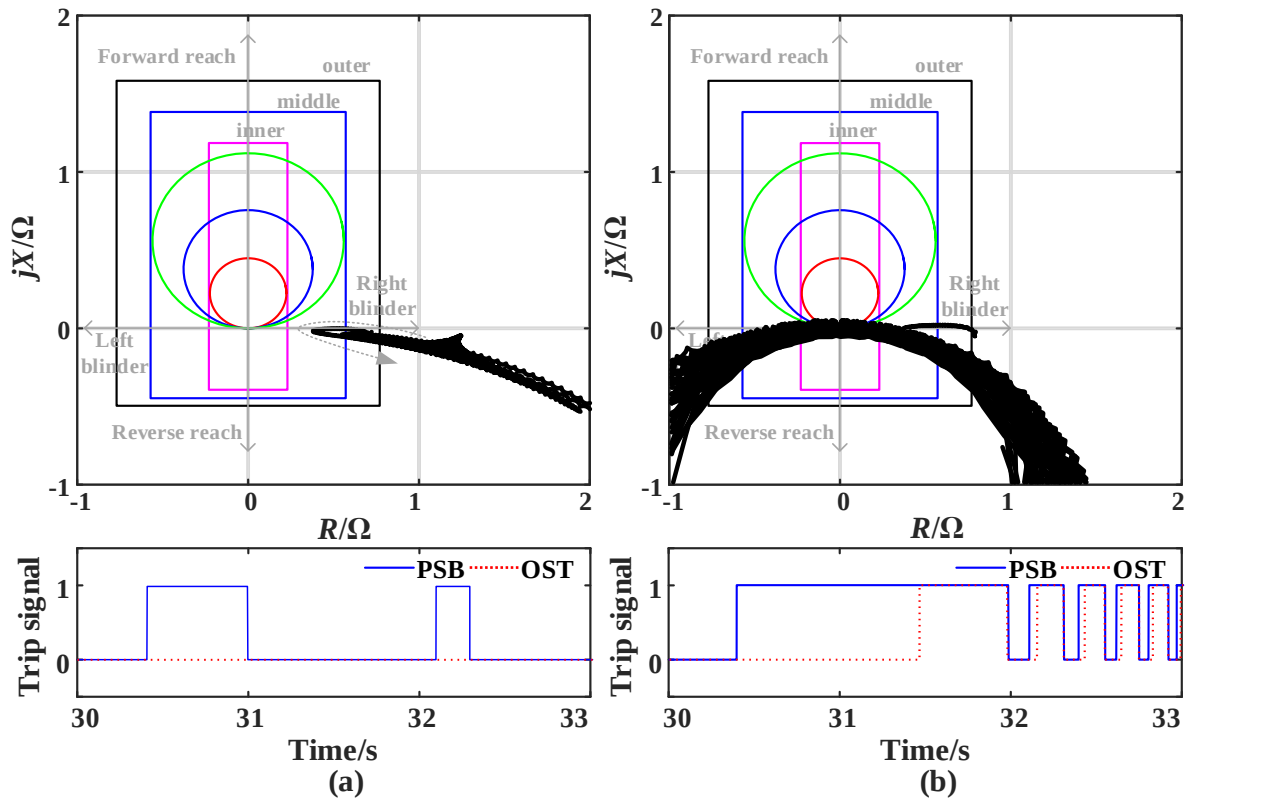

Fig. 15. Impedance trajectories and power swing detection results: (a) stable power swing, (b) unstable power swing.

alternate between the current control mode and the voltage control mode. This leads to a unique swing characteristic for inertial GFM-VSC systems. In such unstable power swings, the impedance trajectory may resemble that of a non-inertial GFM-VSC system during the current control mode and that of an SG-based system during the voltage control mode.

To give a detailed comparison, the impedance trajectories of non-inertial and inertial GFM-VSC systems from Section III-C and Section IV-B are shown in Fig. 13. In the inertial GFM-VSC system with LOS, the impedance trajectory may repeatedly form partial circles during current limitation mode but diverge from the circle in voltage control mode.

In general, in the GFM-VSC system (both non-inertial and inertial), the ***physical angle*** $\delta_S$ can not reflect the swing characteristic of the VSC, and the unstable swing is related to the divergence of the ***control-dependent angle*** $\delta_{PSC}$. Yet, the divergence of $\delta_{PSC}$ does not necessarily correlate to the divergence of the physical power angle $\delta_S$, as shown in Fig. 9. Hence, the traditional power swing detection method that predicts the increase of $\delta_S$ based on the movement of the impedance trajectory might not be effective for GFM-VSC systems and may even lead to maloperations.

Based on (11) and aforementioned analysis, it is possible to quantify the design of the reverse blinder, i.e., it should be lower than $X_g$-$V_g/I_m$ ($V_g/I_m$ is equal to 1/1.2=0.83 p.u.), and provide guidelines for power swing detection in GFM-VSC systems. It also verifies that the traditional impedance-based power swing detection methods for SG-based systems may still operate properly for GFM-VSC under weak grid conditions, but may malfunction as grid stiffness increases.

## VI. Single-Machine Case Studies

### *A. Model Descriptions*

To analyze the control dynamic impact of both the non-inertial and the inertial GFM-VSC systems on traditional impedance-based power swing detection schemes, various time-domain case studies are conducted in the PSCAD/EMTDC platform. For comparison, the power swing characteristics of SG-based systems are also studied with the system depicted in Fig. 1. The line parameters of both the SG and the VSC systems are referred to in Tables I-III.

### *B. Case I: Power Swing in SG-based Systems*

As aforementioned, the power swing characteristics in an SG-based system are characterized by the ***physical angle***. In this scenario, stable and unstable power swings are studied based on the benchmark system depicted in Fig. 1.

A balanced three-phase-ground fault is set at $t$=30s at $X_{g2}$ near the relay $RL_3$. To compare the differences between the stable and unstable power swings, the fault is cleared after 0.26s and 0.27s, corresponding to FCT smaller and larger than the CCT, respectively. The post-fault power swing is observed by $RL_1$. The simulation results and impedance trajectories for stable and unstable power swings in SG-based systems are displayed in Fig. 14 and Fig. 15, respectively.

As shown in Fig. 14 (a), when the FCT is less than the CCT, the power swing is stable and eventually dampens after the fault is cleared, with a swing frequency of approximately 0.95 Hz. According to Fig. 15 (a), the trajectory will travel through the outer and middle blinders twice at $t$=30.32s (with a time interval $\Delta T_{PSB}$ =0.65s) and $t$=32.14s (with a time interval $\Delta T_{PSB}$=0.12s). Thus, the PSB function can operate correctly, The impedance trajectory does not cross the inner blinder, indicating that the OST function is not triggered.

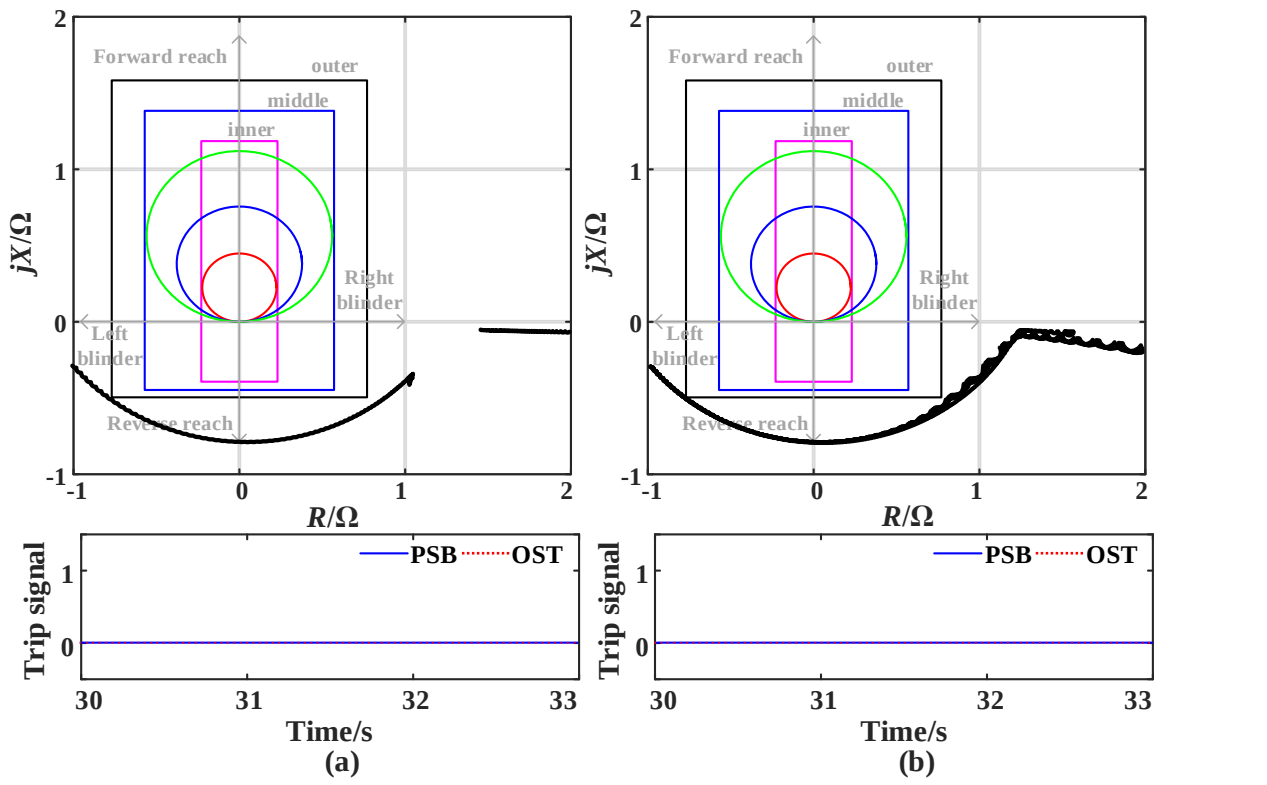

Fig. 16. Impedance trajectory and power swing detection trip signal with conventional power swing detection method in: (a) non-inertial GFM-VSC system, (b) inertial GFM-VSC system.

Conversely, in Fig. 14 (b), when the FCT exceeds the CCT, the post-fault power swing becomes unstable, and the ***physical angle*** ($\delta_S$) may increase larger than 180° and diverge. In Fig. 15 (b), the impedance trajectory crosses the inner blinder, triggering the PSB and OST functions as expected.

Overall, in a conventional SG-based system, stable and unstable power swings can be reliably detected by traditional impedance-based power swing detection methods.

*C. Case II: Power Swing in GFM-VSC Systems*

*(a) With the same settings of SG-based systems*

This scenario investigates the power swing characteristics in both non-inertial and inertial GFM-VSC systems, with the main parameters referred to in Tables II and III, respectively.

The impedance trajectories are shown in Fig. 16. Given the same settings of PSB and OST functions with SG-based systems in Table IV, the impedance trajectory may form part of a circle, as described by (11). The power swings during the current limitation mode are characterized by the ***control-dependent angle*** $\delta_{PSC}$, rather than the ***physical angle*** $\delta_S$. As the impedance trajectory in both the non-inertial and inertial GFM-VSC systems may not cross any blinders, the conventional impedance-based PSB and OST functions can not reliably detect power swings in either non-inertial or inertial GFM-VSC systems.

*(b) With the settings of extended reverse blinder*

Fig. 16 demonstrates that the reverse blinder settings play a crucial role in power swing detection in GFM-VSC systems. In this scenario, all reverse blinders are set to be lower than $X_g$-$V_g$/$I_m$ referred to (11), with outer, middle and inner reverse blinders extended to 0.95Ω, 0.90Ω, and 0.85Ω. Simulation results for power swing detection are presented in Fig. 17.

It can be found that, with the extended reverse blinders, the impedance trajectory successfully crosses all three right blinders in both non-inertial and inertial systems. As shown in Fig. 17 (a), for the non-inertial GFM-VSC system, the time interval for crossing outer and middle blinders is $\Delta T_{non}$=30.527-30.448 =0.079s, while in the inertial GFM-VSC system, the time interval is $\Delta T_{in}$=31.271-31.120 =0.151s, as shown in Fig. 17 (b). Both time intervals are larger than the reference $\Delta T_{PSB}$=0.03s, confirming that power swings can be detected with extended reverse blinders. Moreover, as the impedance trajectories cross the inner right blinder in both non-inertial and inertial GFM-VSC systems, the OST can be triggered in both cases.

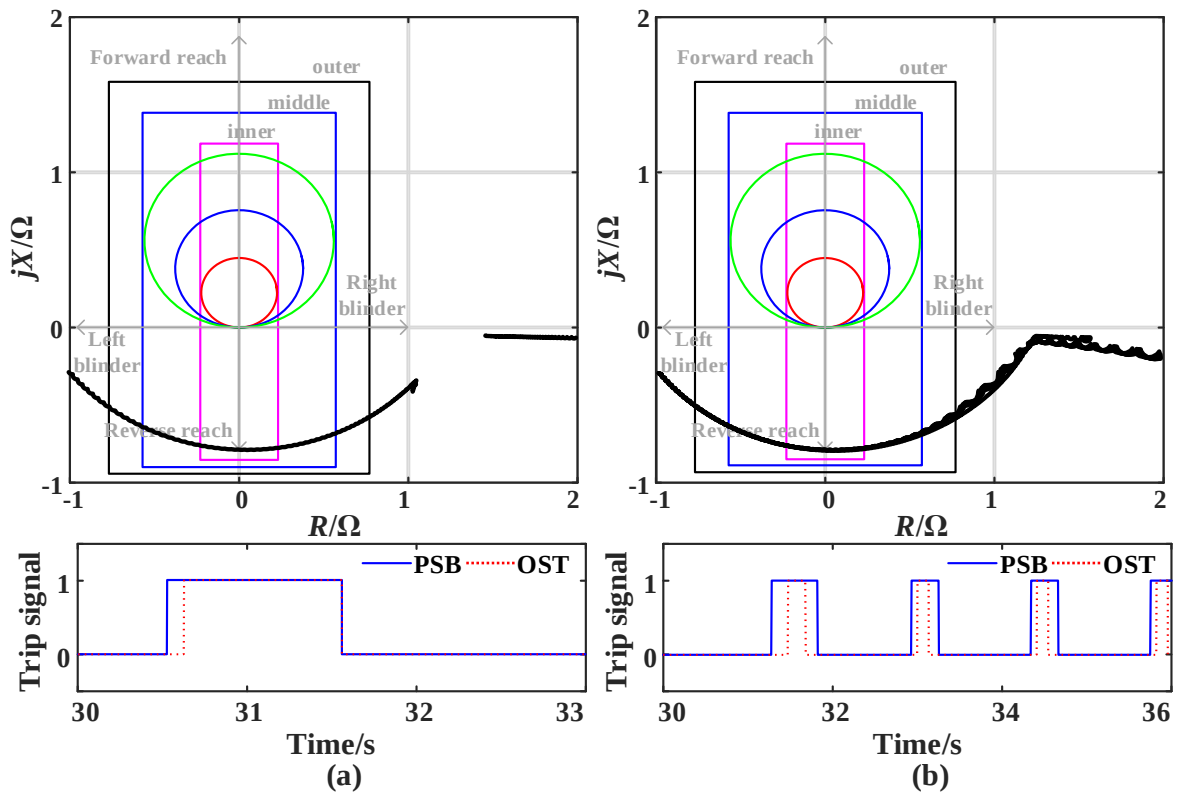

Fig. 17. Impedance trajectory and power swing detection trip signal with extended reverse blinders in: (a) non-inertial GFM-VSC system, (b) inertial GFM-VSC system.

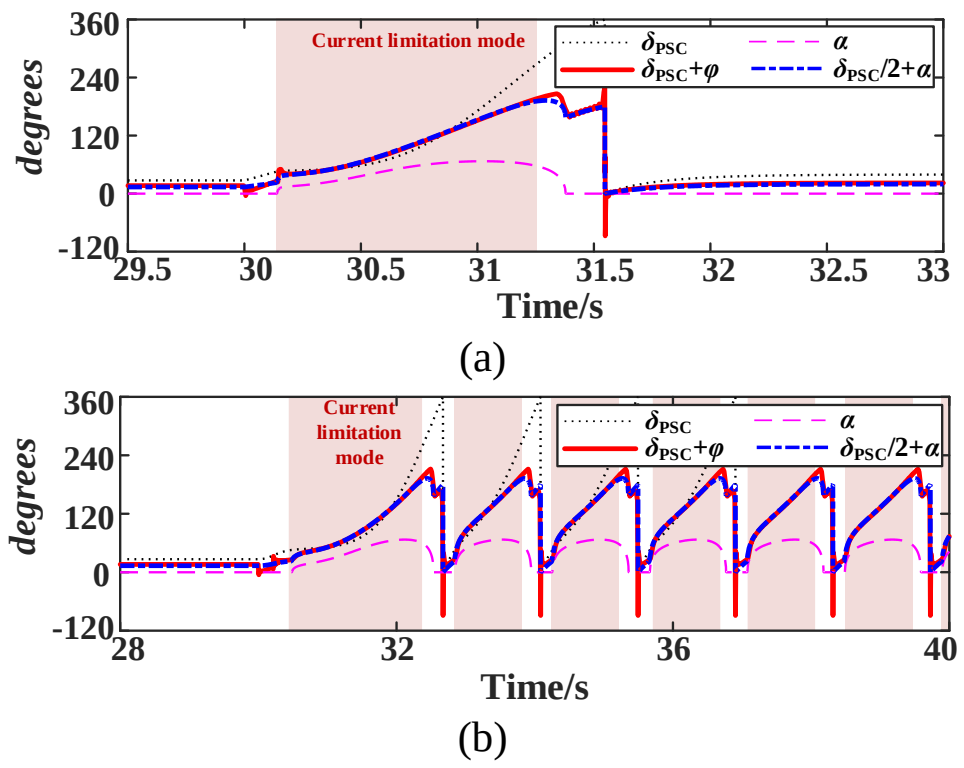

Fig. 18. Control-dependent angle curves during power swings in: (a) non-inertial GFM-VSC system, and (b) inertial GFM-VSC system with LOS.

Based on the simulation results, the proposed quantitative guidelines of the reverse blinder can help to detect the impedance trajectories, and support the impedance-based power swing detection method operating correctly. It should be noted that the activations of OST functions are technically correct for both GFM-VSC systems, but the activation is not preferred in the non-inertial GFM-VSC system, as the VSC can still synchronize with the grid, as illustrated in Fig. 17 (b).

*D. Case III: Impact of the GFM Control-dependent Angle*

In this scenario, the analysis presented in Section V is verified. Focusing on the power swing of the GFM-VSC system as shown in Section VI-B and Fig. 16. The dynamics of the ***physical angle*** $\delta_{PCC}$, and the ***control-dependent angle,*** i.e., $\delta_{PSC}$ in non-inertial and inertial, are depicted in Fig. 7 and Fig. 9, respectively. It can be found that the impedance trajectories in both non-inertial and inertial GFM-VSC systems are characterized by the ***control-dependent angle*** $\delta_{PSC}$:

- In the non-inertial GFM-VSC system, the system can still synchronize with the grid after $\delta_{PSC}$ increases larger than 360 degrees. Conversely, during the power swing, the ***physical angle*** $\delta_{PCC}$ fluctuates within a range of approximately -55 to +50 degrees.
- In the inertial GFM-VSC system, LOS occurs with $\delta_{PSC}$ increasing larger than 360 degrees and finally diverging. In contrast, the ***physical angle*** $\delta_{PCC}$ does not diverge but remains within the same range of -55 to +54 degrees during the LOS period.

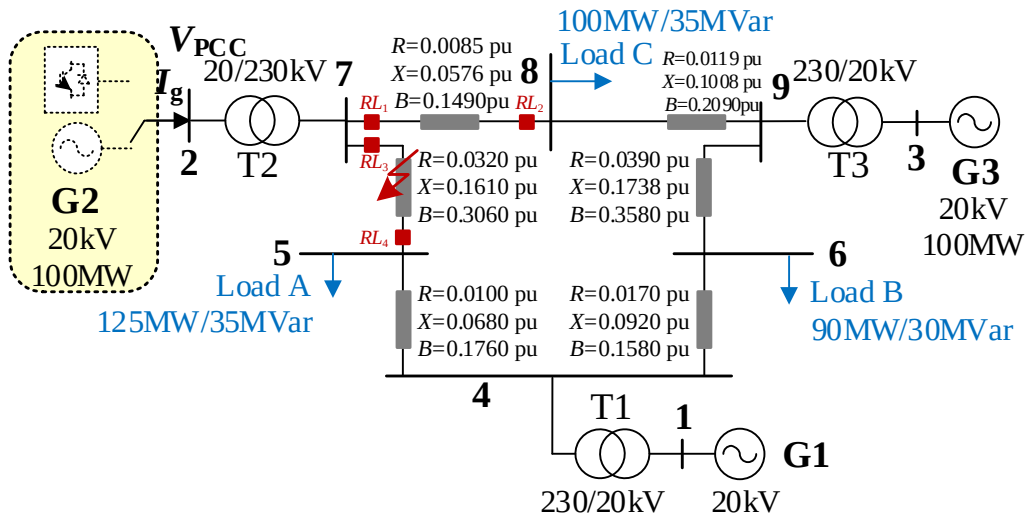

Fig. 19. Modified IEEE 9 bus test system.

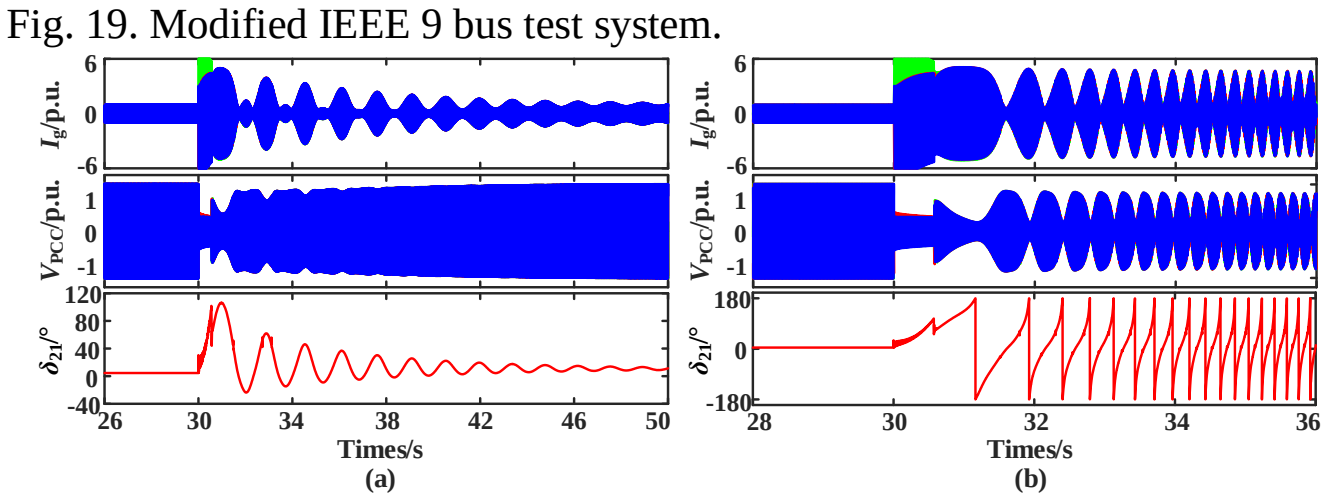

Fig. 20. Simulation results of SG-based systems in multi-machine case:(a) stable power swing, (b) unstable power swing.

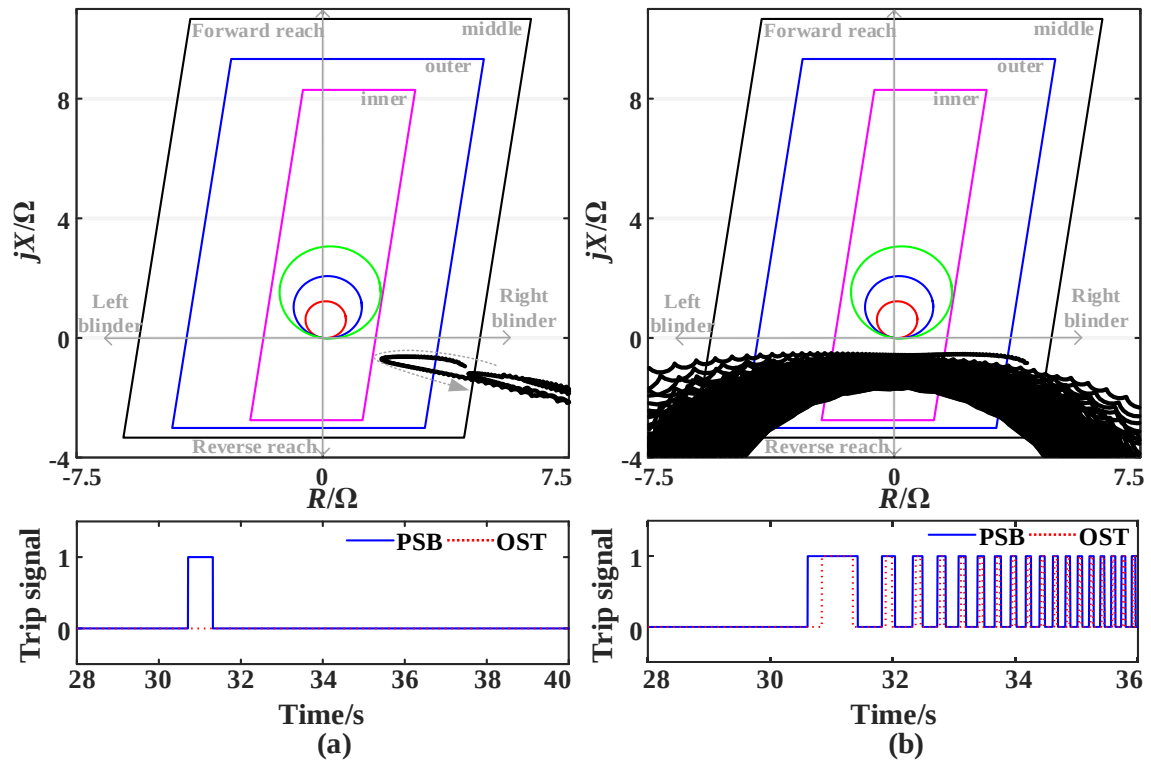

Fig. 21. Impedance trajectories of SG-based systems in the multi-machine case:(a) stable power swing, (b) unstable power swing.

To further verify the impact of the ***control-dependent angle*** on the impedance trajectory, the simulation results of different angle curves are shown in Fig. 18. It can be found that the angle $\alpha$ is consistently greater than zero, as previously analyzed. Additionally, the angle sum $\delta_{PSC}+\varphi$ is always equal to $\delta_{PSC}/2+\alpha$ under the current limitation mode, in both non-inertial and inertial GFM-VSC systems. These findings are consistent with the theoretical analysis in Section V, confirming the validity of the results.

## VII. Multi-Machine Case Studies

### *A. Model Descriptions*

To further investigate the GFM control impact on the impedance-based power swing detection method, the modified IEEE 9-bus system shown in Fig. 19 is employed as the test system [26] [27]. In this system, Generator G1 is replaced by a GFM-VSC connected system, while its rated voltage and active power are settled as the same (i.e., 20kV/100MW). G2 and G3 are SGs. The rated voltage and active power of the transmission system are 230kV/100MW. The main parameters are shown in Fig. 19, with more details referred to [26] [27]. The impedance trajectory is measured at RL1.

### *B. Case I: Power Swing Characteristics in SG-based Systems*

To study both stable and unstable power swing conditions with an SG connection to Bus 2, a three-phase-ground fault is applied at $t$=30s on transmission line L7-5, near Bus 7. For the stable power swing, the fault is cleared by opening RL3 and RL4 after 0.56s. Meanwhile, the fault is cleared after 0.57s for an unstable power swing. Simulation results of three-phase voltage, current at Bus 2, and the phase angle difference between Bus 1 and Bus 2 are depicted in Fig. 20. The impedance trajectory measured by RL1 is shown in Fig. 21.

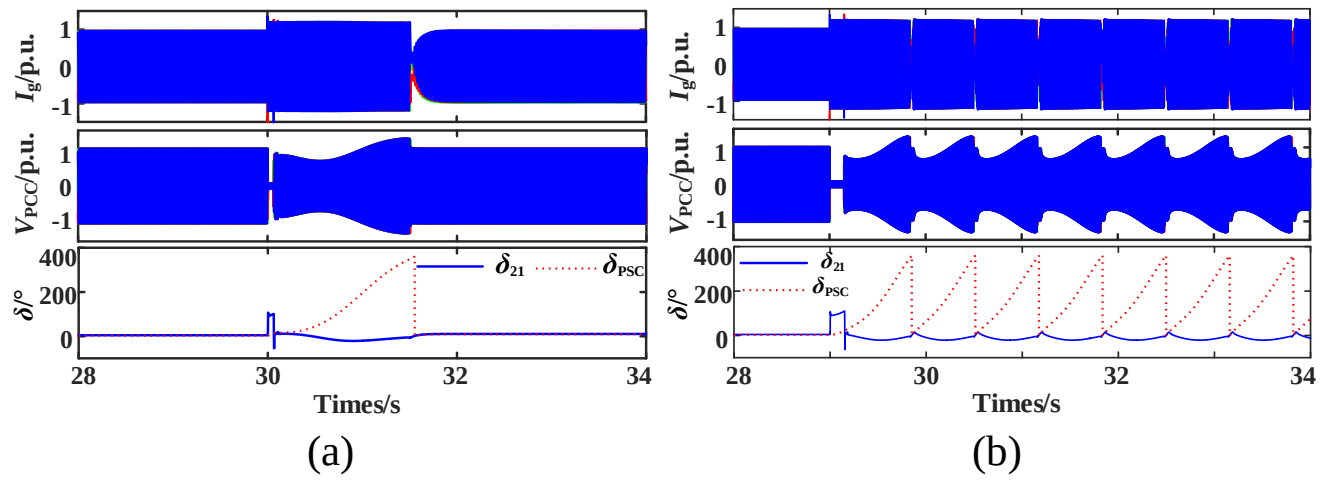

Fig. 22. Simulation results of the multi-machine case with the connection of: (a) the non-inertial GFM-VSC system, and (b) the inertial GFM-VSC system.

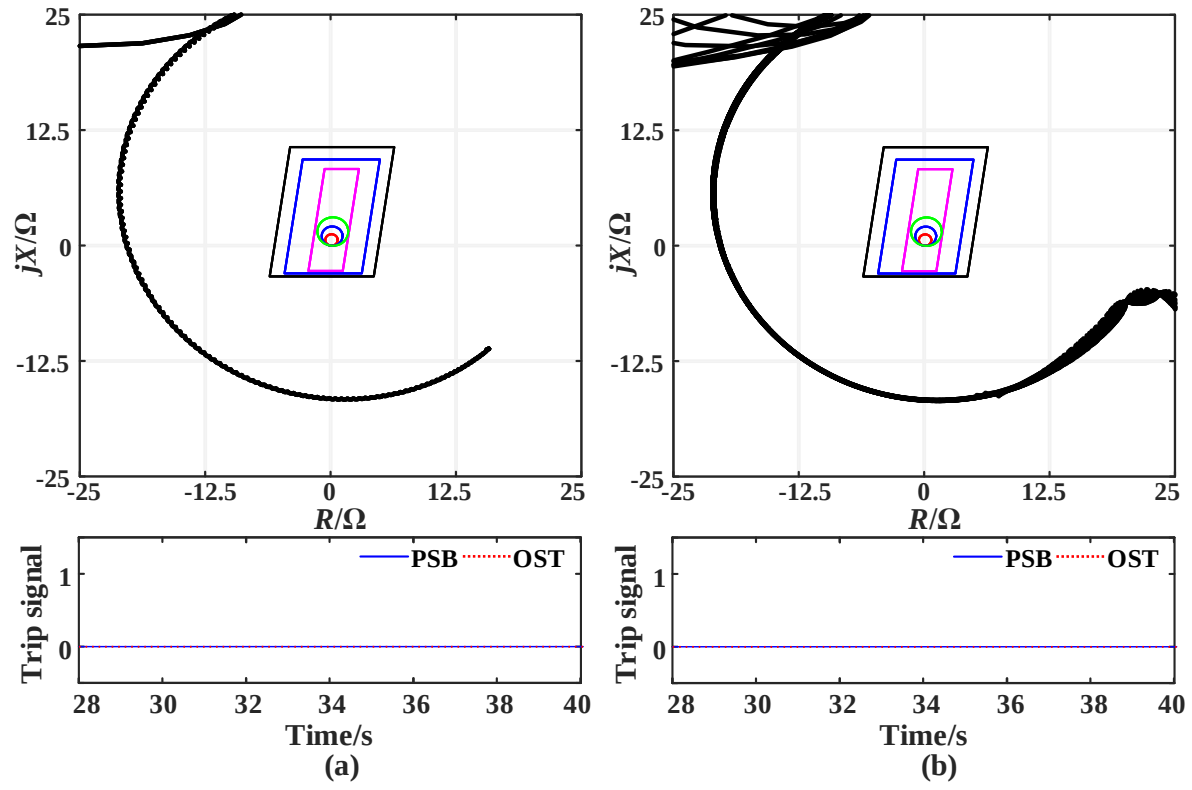

Fig. 23. Impedance trajectories of the multi-machine case with connection of: (a) the non-inertial GFM-VSC system, and (b) the inertial GFM-VSC system.

For the stable power swing, with the SG connected to Bus 2, there is a significant oscillation in the current after the fault. The power swing finally stabilizes with a swing frequency of approximately 0.95 Hz, as shown in Fig. 20 (a). Additionally, the voltage phase angle difference between Bus 2 and Bus 1 ($\delta_{21}$), damps to around 20 degrees after reaching a maximum of 100 degrees. Furthermore, Fig. 21 (a) demonstrates that, with the conventional impedance-based power swing detection method, this stable power swing is detected at $t$=30.8s when the impedance trajectory crosses the middle blinders, triggering the PSB function. Since this is a stable power swing, and the impedance trajectory does not cross the inner blinder, the OST function is not triggered.

For the unstable power swing scenario, as shown in Fig. 20 (b), with the FCT of 0.57s, the power swing becomes unstable after the fault is cleared, with a swing frequency of approximately 5.2 Hz. In this case, $\delta_{21}$ diverges, indicating an unstable power swing. Furthermore, as seen in Fig. 21 (b), the PSB and OST functions are initially triggered at $t$=30.5s and $t$=30.8s, and the efficacy of conventional impedance-based PSB and OST functions in SG-based systems is validated.

### *C. Case II: Powert Swing Characteristics in Non-inertial and Inertial GFM-VSC Systems*

In this section, the efficacy of the impedance-based PSB and OST functions is investigated with the VSC connected at Bus 2 system in the test system depicted in Fig. 19, evaluated for both non-inertial and inertial GFM-VSC systems.

*a) Non-inertial GFM-VSC systems:* In this scenario, a three-phase-to-ground fault occurred at $t$=30s on transmission line L7-5, near Bus 7, and cleared after 0.1s. Simulation results are shown in Fig. 22 (a) and Fig. 23 (a).

It can be found from Fig. 22 (a) that, the ***physical angle*** $\delta_{21}$ may maintain between -21 to 11 degrees, while the ***control-dependent angle*** $\delta_{PSC}$ may increase larger than 360 degrees and re-synchronize with the grid. Besides, with the conventional power swing detection method, the impedance trajectory is characterized by the control-dependent angle $\delta_{PSC}$, and may not cross any of the blinders, meaning that the PSB and OST functions may not be triggered during the re-synchronizing process as shown in Fig. 23 (a). This highlights the importance of reverse blinder settings for swing detection.

*b) Inertial GFM-VSC systems:* In this scenario, the inertial GFM-VSC system is connected to Bus 2, while the other parameters remain the same as the non-inertial GFM-VSC connection scenario. Simulation results are shown in Fig. 22 (b) and Fig. 23 (b). It can be observed from Fig. 22 (b) that $\delta_{21}$ can maintain between -21 to 16 degrees, while $\delta_{PSC}$ may increase and finally diverge. Besides, in Fig. 23 (b) with the conventional impedance-based swing detection method, the trajectory does not cross any blinders, and the PSB and OST functions may also not be triggered during the LOS situation.

## VIII. Conclusion

This study systematically investigates the power swing characteristics of GFM-VSC systems under current limitation mode, focusing on how control dynamics affect traditional impedance-based power swing detection methods. The results demonstrate that in GFM-VSC systems under current limitation mode, power swings are characterized by the control-dependent angle $\delta_{PSC}$, in contrast to the physical angle $\delta_S$ in SG-based systems. This study, for the first time, reveals that the traditional impedance-based power swing detection methods for SG-based systems may still operate properly for GFM-VSC under weak grid conditions, but may malfunction as grid stiffness increases. Additionally, a quantitative design guideline for reverse blinders, which extends them to be lower than $X_g$-$V_g/I_m$, is proposed and verified to be feasible when applied in GFM-VSC systems. The proposed guideline can ensure the accurate tracking of impedance trajectories. The impedance-based detection method was used as a benchmark given its widespread application. Future research will aim to enhance both impedance-based and alternative power swing detection methods in GFM-VSC systems to further improve overall system reliability.